\documentstyle[11pt]{article}
\newcommand{\be}{\begin{equation}}
\newcommand{\ee}{\end{equation}}
\newcommand{\ba}{\begin{eqnarray}}
\newcommand{\ea}{\end{eqnarray}}

\begin{document}
\hbadness=10000
\setcounter{page}{1}

\title{\vspace{-3.0cm} \hspace{0.0cm} \hspace*{\fill} \\[-5.5ex]
\hspace*{\fill}{\normalsize LA-UR-96-1615} \\*[1.5ex]
{\huge \bf The Effects of Bose-Condensates
on Single Inclusive Spectra and Bose-Einstein Correlations} 
}

\author{
U. Ornik${}^1$\thanks{E. Mail: ornik@warp.soultek.de}{\ },
M. Pl\"umer${}^2$\thanks{E. Mail: pluemer@Mailer.Uni-Marburg.DE}{\ }, 
B.R. Schlei${}^3$\thanks{E. Mail: schlei@t2.LANL.gov}{\ },  
D. Strottman${}^3$\thanks{E. Mail: dds@LANL.gov}{\ } and
R.M. Weiner${}^2$\thanks{E. Mail: weiner@Mailer.Uni-Marburg.DE}\\[1.5ex]
{\it ${}^1$SoulTek Internet Services, Software Center 5, Marburg, Germany}\\
{\it ${}^2$Physics Department, Univ. of Marburg, Marburg, Germany}\\
{\it ${}^3$Theoretical Division, Los Alamos National Laboratory, 
Los Alamos, NM 87545}
}

\maketitle

\vspace{0.5cm}

\begin{abstract}
The implications of the formation of a Bose condensate on one- and 
two-particle spectra are studied for ultrarelativistic nucleus-nucleus 
collisions in the framework of a hydrodynamic description. It is found that 
single particle spectra are considerably enhanced at low momenta. The 
Bose-Einstein correlation function has an intercept below two. For pion pairs 
in the central region a two-component structure may appear in the correlation 
function, which is different from that found in quantum optics. The chaoticity 
parameter is strongly momentum dependent.
\end{abstract}

\newpage

\section{Introduction}

In current ultrarelativistic heavy ion collision experiments at 
the SPS (e.g., $Pb+Pb$ at $E_{beam}=160$ AGeV) secondary particles are 
formed at high number densities in rapidity space \cite{pbpb}. 
In future experiments at RHIC and 
the LHC one expects to obtain even higher multiplicities on the order
of a few thousand particles per unit rapidity. If local thermal (but 
not chemical) equilibrium is established and the number densities are 
sufficiently large, the pions may accumulate in their ground state and 
a Bose condensate may be formed\footnote{Recently a different type of 
pion condensate, the disordered chiral condensate, has received a lot 
of attention in the literature (for a review of recent results, 
see Ref. \cite{dcc}). It has been argued \cite{kogan} that
such a condensate would lead to the creation of {\em squeezed},
i.e., two particle coherent states. The possible effects of such
squeezed coherent states on  Bose-Einstein correlations have been studied 
in some detail \cite{dcc1}. 

In what follows, however, we shall restrict ourselves to the case of a 
conventional Bose condensate which is formed when the pion chemical potential 
becomes equal to the pion mass and to the associated one-particle coherent 
states.}. A specific scenario for the formation of a Bose condensate, namely, 
the decay of short-lived resonances, was discussed in Ref. \cite{dan} where 
conditions necessary for the formation of a Bose condensate in a heavy ion 
collision were investigated. In Ref. \cite{dan} it was found that if a pionic 
Bose condensate is formed at any stage of the collision, it can be expected 
to survive until pions decouple from the dense matter, and thus it can affect 
the spectra and correlations of final state pions.

In the present paper we investigate the influence of such a condensate
on the single inclusive cross section and on the second order correlation 
function of identically charged pions (Bose-Einstein correlations BEC) in 
hadronic reactions for {\em expanding} sources.

The proposal to use BEC for the detection of condensates was made a long time 
ago \cite{Fowler}. In that reference the influence of the condensate on the 
form and the intercept of the second and third order correlation function was
studied, with particular emphasis on the possible role of the strong final 
state interactions, which, contrary to more phenomenological approaches used 
afterwards, were studied using a quantum statistical Landau-Ginsburg type 
method.  

It was found in \cite{Fowler} that the interaction influences the shape of the
correlation function but not the value of its maximum (intercept). On the other
hand the amount of condensate influences both the intercept and the form of the
correlation function. In particular in analogy with quantum optics, a typical 
two component structure in the correlation function was found: while for purely
chaotic fields the second order correlation function has only one term which 
depends on the rapidity difference, for a superposition of coherent and chaotic
fields there are two such terms which differ in a well defined manner and which
depend both on the same correlator (cf. eq.(\ref{eq:c2cond0}) below). No 
dependence on the total rapidity of the pair appears in this approach because 
of the assumed boost invariance and of the fact that in the one dimensional 
treatment (in rapidity space) used in \cite{Fowler} no allowance for expansion 
was made.  

The present paper uses a hydrodynamical approach which differs  
significantly from the previous approach in that it considers explicitly the 
expansion of the source and thus the correlation between position and momenta 
of produced particles. Such an approach is certainly more adequate for heavy 
ion reactions and appears desirable at this stage particularly because of the 
possibility mentioned above that a condensate might be produced after the 
expansion of the system during the freeze-out process. Furthermore, the 
momentum dependence of the chaoticity can easily be taken into account. Last 
but not least, in the present paper we exploit explicitly the possibility 
offered by quantum statistics to treat simultaneously single inclusive cross 
sections and BEC and to correlate the possible effects of a condensate in
these observables, which is of considerable interest for future comparisons 
with experiment.

The remainder of this paper is organized as follows. In section 2 general 
expressions are derived for the single particle spectrum and two-particle 
correlation function of pions emitted from a hydrodynamically expanding fluid 
with a superfluid component. In sections 3 and 4 the formalism is applied to 
the case of a spherical expansion and a longitudinal scaling expansion, 
respectively. Numerical results for both models, obtained for specific 
parametrizations of the freeze-out hypersurface, are presented in section 5. 
Finally, the main results of the paper are summarized in section 6.

\section{Basic formalism}

Applying the current formalism \cite{shuryak,podgor,gkw,apw} we write the 
amplitude for the emission of a pion of four-momentum $k=(E,\vec{k})$ as a 
superposition of contributions from the thermal excitations and from the 
condensate, 
\be
J(k) = J_{th}(k) + J_{co}(k) \:,
\label{eq:current} 
\ee
with
\be
J_{th}(k) = \sum_l J_l^{th}(k)
= \sum_l \displaystyle{e^{i \phi_l^{th}}
e^{i k x_l}} j_l^{th}(k) \:,
\label{eq:thcurrent}
\ee
\be
J_{co}(k) = \sum_n J_n^{co}(k)
= \sum_n \displaystyle{e^{i \phi_n^{co}}
e^{i k x_n}} j_n^{co}(k) \:,
\label{eq:cocurrent}
\ee
where the labels ``{\it th}'' and ``{\it co}'' indicate the thermal and the 
condensate component of the distribution. The indices $l$ and $n$ label source 
elements centered at space-time point $x_l$ and $x_n$, respectively, and 
$\exp(i\phi_l^{th})$ and $\exp(i\phi_n^{co})$ are the corresponding fluctuating
phase factors. 

For the thermal component, two phases that characterize the emission from two 
different source elements,  $\phi_l^{th}$ and $\phi_{l'}^{th}$ with $l\neq l'$,
are taken to be uncorrelated. Likewise, we assume that there exists no 
correlation between the phases related to emission from the thermal component 
and from the condensate, $\phi_l^{th}$ and $\phi_n^{co}$. If, and to what 
degree, the phase factors for emission from the condensate are correlated for 
two different source elements depends on the details of the formation and 
evolution of the condensate. There are two extreme cases: 
\begin{description}
\item[(a)] The phases $\phi_n^{co}$ are completely correlated, i.e., the 
differences $\phi_n^{co}-\phi_{n'}^{co}$ do not fluctuate. In the of case an 
expanding source this would imply that the condensate must have been formed at 
an early stage when the hadronic matter was concentrated in a space-time volume
sufficiently small for this kind of global phase coherence to be established. 
This corresponds to the presence of an expanding superfluid with a globally 
coherent phase.   
\item[(b)] The phases $\phi_n^{co}$ are completely uncorrelated. This is what 
one would expect if the condensate is formed at a late stage when the process 
of condensate formation (caused, e.g., by the decay of short-lived resonances)
occurs independently in different fluid cells. 
\end{description}
In the following we shall consider only the case (b), i.e., we assume that the
Bose condensate is not phase correlated in different fluid cells. This 
situation appears as more realistic than case (a) although the formation of a 
global superfluid as envisaged in (a) is in itself also of interest and might
deserve further study. 

The single and double inclusive momentum distribution are 
\be
E\frac{d^3 N}{d^3 k} = \langle 
J^\star(k) J(k) \rangle \:,
\label{eq:singlein}
\ee
and
\be
E_1 E_2 \frac{d^6 N}{d^3 k_1 d^3 k_2} = 
\langle J^\star(k_1) J^\star(k_2)
J(k_2) J(k_1) \rangle \:,
\label{eq:doublein}
\ee
where the averaging is performed over the space-time positions $x_l$, $x_n$ and
the phases $\phi_l^{th}$, $\phi_n^{co}$. 

It is useful to introduce the following correlators: 
\be
D^{th}(k_1,k_2) \equiv \bigg\langle
\sum_l \left( J_l^{th}(k_1) \right)^\star
J_l^{th}(k_2) \bigg\rangle \:,
\label{eq:dkkth}
\ee
\be
D^{co}(k_1,k_2) \equiv \bigg\langle
\sum_n \left( J_n^{co}(k_1) \right)^\star
J_n^{co}(k_2) \bigg\rangle \:,
\label{eq:dkkco}
\ee
\be
G^{th}(k_1,k_2) \equiv \bigg\langle
\sum_l \left|J_l^{th}(k_1)\right|^2 
\left|J_l^{th}(k_2)\right|^2 \bigg\rangle \:,
\label{eq:gkkth}
\ee
\be
G^{co}(k_1,k_2) \equiv \bigg\langle
\sum_n \left|J_n^{co}(k_1)\right|^2 
\left|J_n^{co}(k_2)\right|^2 \bigg\rangle \:.
\label{eq:gkkco}
\ee
The particle spectrum may then be expressed as the sum of a thermal and a 
condensate contribution, 
\be
E\frac{d^3 N}{d^3 k} = 
E\frac{d^3 N}{d^3 k}\bigg|_{th} + 
E\frac{d^3 N}{d^3 k}\bigg|_{co} \:,
\label{eq:singzerl} 
\ee
with
\be
E\frac{d^3 N}{d^3 k}\bigg|_{th} = D^{th}(k,k)\:,
\label{eq:thsing}
\ee
and 
\be
E\frac{d^3 N}{d^3 k}\bigg|_{co} = D^{co}(k,k) \:,
\label{eq:cosing}
\ee
and the two-particle inclusive distributions may be written as
\ba
\langle J^\star(k_1) J^\star(k_2)
J(k_2) J(k_1) \rangle &=&
\left|D^{th}(k_1,k_2)+D^{co}(k_1,k_2)\right|^2 \nonumber\\
 & & - G^{th}(k_1,k_2) - G^{co}(k_1,k_2).
\label{eq:wick}
\ea
The two-particle correlation function then takes the form
\ba
C_2(\vec{k}_1,\vec{k}_2) &=& 
\frac{\langle J^\star(k_1) J^\star(k_2)
J(k_2) J(k_1) \rangle}
{\langle J^\star(k_1) J(k_1) \rangle
\langle J^\star(k_2) J(k_2) \rangle}\nonumber\\
&=& 1 + \frac{\left|D^{th}(k_1,k_2)+ D^{co}(k_1,k_2)
  \right|^2 - G^{th}(k_1,k_2) - G^{co}(k_1,k_2)}
{(D^{th}(k_1,k_1)+ D^{co}(k_1,k_1))
(D^{th}(k_2,k_2)+ D^{co}(k_2,k_2))
}\:.
\label{eq:c2g}
\ea

We first consider the thermal correlators $D^{th}(k_1,k_2)$ and 
$G^{th}(k_1,k_2)$ to further evaluate these expressions. Let $L$ be the number 
of source elements that contribute to particles of momenta $\vec{k}_1$ and 
$\vec{k}_2$. In the limit of large $L$ the term $G^{th}(k_1,k_2) \propto L$ can
be neglected compared to the term $|D^{th}(k_1,k_2)|^2 \propto L^2$ in eq. 
(\ref{eq:wick}). In the case of hydrodynamics, the sum over source elements 
translates into an integral over the freeze-out hypersurface. Introducing the 
average and the relative four-momentum of the pair, $K \equiv 
\textstyle{\frac{1}{2}}(k_1 + k_2)$ and $q \equiv k_1 - k_2$, respectively,
one has
\be
D^{th}(k_1,k_2) = \int d^4 x\ 
g^{th}(x,K)\  \displaystyle{e^{i q \: x}}\:,
\label{eq:dk1k2}
\ee
with the thermal source function for pions \cite{bernd1}
\be
g^{th}(x,k) = \frac{g_\pi}{(2\pi)^3}
\int_\Sigma \frac{k \: d\sigma(x^\prime) \:
\delta^4(x - x^\prime)}
{\displaystyle{\exp\left[\frac{
k \: u(x^\prime)-\mu_\pi(x^\prime)}
{T_f(x^\prime)}\right]}-1}\:.
\label{eq:gth}
\ee
In eq. (\ref{eq:gth}) $d\sigma(x)$, $u(x)$ and $T_f(x)$ are the volume element 
of the freeze-out hypersurface $\Sigma$, the 4-velocity of the fluid and the 
freeze-out temperature at space-time point $x$, respectively. The factor 
$g_\pi$ denotes the degeneracy factor of the pions while $\mu_\pi(x)$ is the 
pionic chemical potential at space-time point $x$. For simplicity, in the 
following sections we shall assume that $\mu_\pi(x) = \mu_\pi = const.$, 
i.e., that a pionic Bose condensate is formed at each space-time point on the 
freeze-out hypersurface.  

Having discussed the terms $D^{th}(k_1,k_2)$ and $G^{th}(k_1,k_2)$ which are
due to the thermal part, we now  proceed to consider the corresponding 
expressions related to the condensate, $D^{co}(k_1,k_2)$ and $G^{co}(k_1,k_2)$.
In what follows each fluid element will be treated as a macroscopic system in 
so far as it will be assumed that the condensate in each fluid cell is 
identified with the lowest momentum state in the rest frame of that fluid cell.
This is in line with the conventional meaning of the concept of a condensate 
which refers to a phase of matter. That is to say, for a fluid cell centered 
around $x_l$ moving with velocity $\vec{u}_l$ we have   
\begin{equation}
|J_l^{co}(k)|^2
\propto \delta^3(\vec{k}- m_\pi \vec{u}_l),
\label{eq:delta}
\end{equation} 
i.e., as emphasized in Ref. \cite{dan}, particles emitted from the condensate 
move with the collective velocity of the fluid. In the models of expanding 
sources which will be discussed in Sections 3 and 4 there is a one-to-one 
relation between the position $x_l$ of the source element and its velocity 
$u_l$. As far as particles emitted from the condensate are concerned, eq. 
(\ref{eq:delta}) then implies that for each momentum $\vec{k}$ only {\it one 
single source element} contributes to the spectrum. Note that nevertheless 
this contribution will be comparable in magnitude to the contributions of the 
thermal excitations since the condensate constitutes a macroscopically occupied
quantum state.  

From eqs. (\ref{eq:dkkco}), (\ref{eq:gkkco})  and 
(\ref{eq:delta}) it follows that  
\begin{equation}
D^{co}(k_1,k_2)\ =\ G^{co}(k_1,k_2)\ =\ 0 \qquad {\rm for} \ k_1 \neq k_2. 
\label{eq:dcogco}
\end{equation}
The contribution of the condensate to the single inclusive spectrum can then be
written as  
\begin{equation}
\frac{d^3 N}{d^3 k}\bigg|_{co}  \ = \ D^{co}(k,k)\nonumber\\
\ = \ \int d^4 x \:g^{co}(x,k)\:
\label{eq:sico}
\end{equation}
with \cite{dan}
\be
g^{co}(x,k) \:=\:
 \int_\Sigma  d\sigma(x^\prime) u(x^\prime)\: n_{co}(x^\prime)\: 
E \: \delta^3(\vec{k}-m_\pi \vec{u}) \:
\delta^4(x - x^\prime)\:.
\label{eq:gco}
\ee
We define the fraction of thermally produced particles of four-momentum $k$, 
i.e., the momentum dependent chaoticity $p(k)$, as
\be
p(k) \equiv 
\frac{\displaystyle{E\frac{d^3 N}{d^3 k}\bigg|_{th}}}
{\displaystyle{E\frac{d^3 N}{d^3 k}}} =
\frac{D^{th}(k,k)}  
{D^{th}(k,k) + D^{co}(k,k)} 
\label{eq:pallg} 
\ee
and the normalized correlator of two thermal currents 
\begin{equation}
d^{th}(k_1,k_2)\  =\  
\frac{D^{th}(k_1,k_2)}  
{\left[D^{th}(k_1,k_1) D^{th}(k_2,k_2)\right]^{1/2}}\: . 
\label{eq:smalld}
\end{equation}
Substituting the expressions  (\ref{eq:dk1k2}), (\ref{eq:gth}) 
and (\ref{eq:sico}) -- (\ref{eq:smalld}) into (\ref{eq:c2g})
we find for the two-particle Bose-Einstein correlation 
function 
\ba
C_2(\vec{k}_1,\vec{k}_2) & = & 1 + 2 \sqrt{p(k_1)p(k_2)(1-p(k_1))(1-p(k_2))}
\ \tilde{\Theta}(|\vec{k_1}-\vec{k_2}|) \ d^{th}(k_1,k_2)\nonumber\\
 & & + \ p(k_1) p(k_2) \ |d^{th}(k_1,k_2)|^2\:, 
\label{eq:c2cond0}
\ea
where 
\be
\tilde{\Theta}(x) \ = \ 
\left\{
\begin{array}
{r@{\quad:\quad}l}
1 & x \leq 0\\
0 & x>0
\end{array}
\right. \:,
\ee
i.e., one has 
\be
C_2(\vec{k}_1,\vec{k}_2) = 1 + p(k_1)\:p(k_2)\:
|d^{th}(k_1,k_2)|^2
\quad {\rm for} \quad k_1 \neq k_2 \: . 
\label{eq:c2cond}
\ee
Note that the ``true'' intercept -- obtained by evaluating (\ref{eq:c2cond0}) 
at $\vec{k_1}=\vec{k_2}$ -- is given by
\begin{equation} 
C_2(\vec{k},\vec{k}) = 1 + 2 p(k) - p^2(k)\:,
\label{eq:truei}
\end{equation}
whereas the value obtained 
by extrapolating (\ref{eq:c2cond}) to $\vec{k}_1=\vec{k}_2$ is  
\begin{equation} 
C_2^{extrapol.}(\vec{k},\vec{k}) = 1 +  p^2(k)\: .
\label{eq:extrapoli}
\end{equation}

The reason for the difference between (\ref{eq:truei}) and (\ref{eq:extrapoli})
is that in the idealized case considered here $D^{co}(\vec{k}_1,\vec{k}_2)$ and
$G^{co}(\vec{k}_1,\vec{k}_2)$ vanish for $\vec{k}_1 \neq \vec{k}_2$ (cf. Eq. 
(\ref{eq:dcogco})), while 
\begin{equation}
G^{co}(\vec{k},\vec{k})\ =\  \left|D^{co}(\vec{k},\vec{k})\right|^2 \ =\ 
\left|J_l(\vec{k})\right|^4 \:,
\label{eq:jl4}
\end{equation}
where $l$ labels the single source elements which contributes to the emission
of pions of momentum $\vec{k}$ from the condensate 
(cf. eqs. (\ref{eq:dkkco},\ref{eq:gkkco},\ref{eq:delta})).

\section{Spherically expanding source}

Let us assume a space-like freeze-out hypersurface parametrized as $t = t(r)$  
where $t=x^0$ is the time coordinate and $r=|\vec{x}|$ the radial coordinate. 
We also assume a radial velocity field $\vec{u} = u(r)\:\vec{e}_r$ with 
non-vanishing gradient, i.e., $(\partial u/ \partial r) \neq 0$. The pionic 
freeze-out temperature is taken to be $T_f = m_\pi = const.$ The volume element
of the 3-dimensional freeze-out hypersurface and the 4-velocity of the 
spherically expanding relativistic fluid are ($\nu=0,1,2,3$)
\be
d\sigma^\nu(x)\:=\:\left(1\:,\:\displaystyle{
\frac{\partial t}{\partial r}}
\vec{e}_r \right)\: r^2\:dr\:\sin\theta\:d\theta\:d\phi\:,
\label{eq:dsigsph}
\ee
\be
u^\nu(x)\:=\:(u^0(r),u(r)\vec{e}_r)
\:,\quad u^0(r)\:=\:\sqrt{1+[u(r)]^2}.
\label{eq:umusph}
\ee

\subsection{Single inclusive momentum distributions}

With the 4-momentum
$k^\nu = (E\:,\:\vec{k}) = (E\:,\:k \cdot \vec{e}_k)$
and $\vec{e}_r\cdot\vec{e}_k = \cos \theta \equiv z$,  
the thermal part of the single inclusive
momentum distribution is given through
\ba
E\frac{d^3 N}{d^3 k}\bigg|_{th}
&=&\frac{g_\pi}{(2\pi)^3} \int_\Sigma
d\sigma^\nu k_\nu \frac{1}{
\exp\left[(u^\nu k_\nu - \mu_\pi)/T_f\right] -1}
\nonumber\\
&=&\frac{g_\pi}{(2\pi)^2}
\int_0^{R_\perp} r^2 dr \int_{-1}^1 dz\:
\frac{E- k z \partial t / \partial r}
{\exp\left[(E u^0(r)-k z u(r) - \mu_\pi)/T_f\right]-1}\:.
\label{eq:sithsph}
\ea     

In eq. (\ref{eq:sithsph}) $E=\sqrt{m_\pi^2+k^2}$ and $R_\perp$ is the radial 
extension of the pion source.

The coherent part of the single inclusive momentum distribution is
\ba
E\frac{d^3 N}{d^3 k}\bigg|_{co}
&=& n_{co} \int_\Sigma d\sigma^\nu u_\nu
E \delta^3 (\vec{k}-m_\pi\vec{u})
\nonumber\\
&=&n_{co} \int_0^{R_\perp} r^2 dr \int_0^\pi \sin\theta \: d\theta
\int_0^{2\pi} d\phi \left(u^0(r)-u(r)\frac{\partial t}
{\partial r}\right)
\nonumber\\
&&\quad\times\:E\:\delta^3(k\vec{e}_k-m_\pi u(r)\vec{e}_r)
\nonumber\\
&=&
\left\{
\begin{array}
{r@{\quad:\quad}l}
n_{co}\frac{E}{m_\pi}
\left(\frac{E}{m_\pi}-\frac{k}{m_\pi}\frac{\partial t}
{\partial r}(r_0) \right)
\left(\frac{r_0}{k}\right)^2 \left(
\frac{\partial u}{\partial r}(r_0) \right)^{-1} & k \leq m_\pi u_{max}\\
0 & k > m_\pi u_{max}
\end{array}
\right. \:,
\label{eq:sicosph}
\ea     

where $r_0=r(u=\frac{k}{m_\pi})$ and $u_{max}$ is the maximal value of the 
velocity at freeze-out. The total momentum distribution is obtained by 
inserting  eqs. (\ref{eq:sithsph}) and (\ref{eq:sicosph}) into eq. 
(\ref{eq:singzerl}).\\

\subsection{Bose-Einstein correlation functions}

For illustration we restrict ourselves to the central momentum region. To be 
specific, we consider the case $\vec{K} \equiv 
\displaystyle{\frac{1}{2}}(\vec{k}_1+\vec{k}_2) = 0$, i.e., $\vec{k}_1 = 
-\vec{k}_2$. The quantity that remains to be calculated in order to construct 
the BEC function is the thermal correlator $D^{th}(k_1,k_2)$. With  $E_1 = E_2$
one has $q^0 \equiv E_1 - E_2 = 0$. Using the notation $\vec{q} \equiv 
\vec{k}_1 - \vec{k_2}$ we obtain $q^\nu x_\nu = - \vec{q} \cdot \vec{x}
= - q r \cos\theta$ and
\ba
D^{th}(k_1,k_2)\bigg|_{\vec{K}=0}&=&
\frac{g_\pi}{(2\pi)^3} \int_\Sigma
\frac{K^\nu d\sigma_\nu e^{i q^\nu x_\nu}}{
\exp\left[(K^\nu u_\nu - \mu_\pi)/T_f \right]-1}
\nonumber\\
&=&\frac{g_\pi}{(2\pi)^3} \int_0^{2\pi} d\phi
\int_0^{\pi} \sin\theta d\theta \int_0^{R_\perp}
r^2 dr\: \frac{E_K e^{-i q r \cos\theta}}
{\exp\left[(E_K u^0(r) - \mu_\pi)/T_f \right]-1}
\nonumber\\
&=&\frac{g_\pi}{(2\pi)^2} \int_0^{R_\perp}
r dr\: \frac{2 E_K \sin(qr)/q}
{\exp\left[(E_K u^0(r) - \mu_\pi)/T_f \right]-1}\:.
\label{eq:d12sph}
\ea     

In section 5 explicit parametrizations for $t(r)$ and $u(r)$ will be used to 
evaluate numerically the above expressions.\\

\section{Longitudinally expanding source}

Here we assume a space-like freeze-out hypersurface parametrized by the 
radius-dependent longitudinal proper-time $\tau(r)$. For the longitudinal 
component of the velocity we take the scaling ansatz $v_{||}=x_{||}/t$. To be 
specific, the 4-volume element of the freeze-out hypersurface and the 
4-velocity of the relativistic fluid are written as 
\be
d\sigma^\nu = (\:\cosh\eta\:,\:\sinh\eta\:,\:
\displaystyle{\frac{\partial \tau}{\partial r}}\vec{e}_r\:)
\:\tau(r)\:d\eta\:r\:dr\:d\phi\:,
\label{eq:dsig4bjo}
\ee
\be
u^\nu = (\:U(r)\cosh\eta\:,\:
U(r)\sinh\eta\:,\:
u_\perp (r)\vec{e}_r\:)\:,\quad 
U(r) = \sqrt{1+[u_\perp (r)]^2}\:,
\label{eq:u4bjo} 
\ee

where $u_\perp(r)$ and $\eta$ are the transverse (radial) component of the 
fluid 4-velocity and the space-time rapidity of the fluid, respectively.\\

\subsection{Single inclusive momentum distributions}

Since our ansatz is invariant under boosts in the longitudinal (beam) direction
we may without loss of generality restrict ourselves to the case of rapidity 
$y=0$. With the 4-momentum $k^\nu = (E\:,\:\vec{k}) = 
(m_\perp\cosh y\:,\:m_\perp\sinh y\:,\:k_\perp\vec{e}_k)$
and $\vec{e}_r\cdot\vec{e}_k \equiv \cos\psi$ the thermal part of the single 
inclusive momentum distribution is given by 
\ba
&&E\frac{d^3 N}{d^3 k}\bigg|_{th}
\:=\:\bigg[\frac{g_\pi}{(2\pi)^3} \int_\Sigma
d\sigma^\nu k_\nu \frac{1}{
\exp\left[(u^\nu k_\nu - \mu_\pi)/T_f\right] -1}\bigg]_{y\:=\:0}
\nonumber\\
&&=\:\frac{g_\pi}{(2\pi)^3}\int_{-\infty}^{+\infty} d\eta\:  
\int_0^{R_\perp} \tau(r) r dr \int_0^{2\pi} d\psi\:
[m_\perp \cosh \eta- k_\perp \cos\psi \:(\partial \tau / \partial r)] 
\nonumber\\
&&\quad \times\:\frac{1}
{\exp\left[(m_\perp \cosh\eta \:U(r)-k_\perp \cos\psi \:u_\perp (r) 
- \mu_\pi)/T_f\right]-1}\:.
\label{eq:sithbjo}
\ea     

In eq. (\ref{eq:sithbjo}) $R_\perp$ is the radial extension of the pion source 
and $m_\perp=\sqrt{m_\pi^2+k_\perp^2}$.

The coherent part of the single inclusive momentum distribution is
\ba
E\frac{d^3 N}{d^3 k}\bigg|_{co}
&=&\bigg[ n_{co} \int_\Sigma d\sigma^\nu u_\nu
E \delta^3 (\vec{k}-m_\pi\vec{u})\bigg]_{y\:=\:0}
\nonumber\\
&=&n_{co} \int_{-\infty}^{+\infty} d\eta \int_0^{R_\perp} \tau(r) r dr 
\int_0^{2\pi} d\phi \left(U(r)-u_\perp (r)\frac{\partial \tau}
{\partial r}\right)
\nonumber\\
&&\quad\times\: m_\perp\:\delta^3(\vec{k}-m_\pi \vec{u})\bigg|_{y\:=\:0}
\nonumber\\
&=&
\left\{
\begin{array}
{r@{\quad:\quad}l}
n_{co}\frac{r_0\:\tau(r_0)}{m_\pi\:k_\perp}
\left(\frac{m_\perp}{m_\pi}-\frac{k_\perp}{m_\pi}\frac{\partial \tau}
{\partial r}(r_0) \right)
\left(\frac{\partial u_\perp}{\partial r}(r_0) \right)^{-1} & k_\perp 
\leq m_\pi u_{max}\\
0 & k_\perp > m_\pi u_{max}
\end{array}
\right. \:,
\label{eq:sicobjo}
\ea     

where $r_0=r(u_\perp=\frac{k_\perp}{m_\pi})$ and $u_{max}$ is the maximal value
of the transverse velocity field at freeze-out. The total momentum distribution
is obtained by inserting (\ref{eq:sithbjo}) and (\ref{eq:sicobjo}) into eq. 
(\ref{eq:singzerl}).\\

\subsection{Bose-Einstein correlation functions}

As in the preceeding section we restrict ourselves to the case $\vec{K} = 
\displaystyle{\frac{1}{2}}(\vec{k}_1+\vec{k}_2) \equiv 0$, i.e., $\vec{k}_1 = 
-\vec{k}_2$. Since we have assumed the source to be invariant under boosts in 
longitudinal direction, the generalization to average pair momenta 
$\vec{K}_\perp=0,K_{||}\neq 0$ is straightforward. We shall consider the 
correlation function in the transverse direction $C_2(q_\perp,\Delta y=0)$, 
i.e., the case $K_\parallel=k_{i\parallel}=y_i=0$ and 
$E_K=E_i=m_{\perp K}=\sqrt{m_\pi^2+k_{i\perp}^2}$ $(i=1,2)$. Then for 
construction of the BEC function the only remaining term to specify is 
$D(k_1,k_2)$.\\
With $q^0 \equiv E_1 - E_2 = 0$ we obtain $q^\nu x_\nu = - \vec{q} \cdot 
\vec{x} \equiv - q_\perp r \cos\psi$ (since $q_{\parallel}=0$) and
\ba
&&D(k_1,k_2)\bigg|_{\vec{K}=0}(\Delta y=0)\:=\:
\frac{g_\pi}{(2\pi)^3} \int_\Sigma
\frac{K^\nu d\sigma_\nu e^{i q^\nu x_\nu}}{
\exp\left[(K^\nu u_\nu - \mu_\pi)/T_f \right]-1}
\nonumber\\
&&=\:\frac{g_\pi}{(2\pi)^3} \int_{-\infty}^{+\infty} d\eta \int_0^{2\pi} d\psi
\int_0^{R_\perp}
r dr\: \frac{\tau(r) m_{\perp K}\: \cosh\eta\: e^{-i q_\perp r \cos\psi}}
{\exp\left[(E_K U(r) - \mu_\pi)/T_f \right]-1}
\nonumber\\
&&=\frac{g_\pi}{(2\pi)^2} \int_{-\infty}^{+\infty} d\eta \int_0^{R_\perp}
r dr\: \frac{\tau(r) m_{\perp K}\: \cosh\eta \:J_0(q_\perp r)}
{\exp\left[(m_{\perp K} U(r) \cosh\eta\ - \mu_\pi)/T_f \right]-1}\:,
\label{eq:d12bjo}
\ea     

with the Bessel function $J_0$.\\

In the next section explicit parametrizations for $\tau(r)$ and $u_\perp(r)$ 
will be given and the results for eqs. (\ref{eq:sithbjo}),(\ref{eq:sicobjo}) 
and (\ref{eq:d12bjo}) will be discussed after their numerical evaluation.\\

\section{Parametrizations and Results}

For the sake of illustration of the above results, we shall consider the case 
of a specific heavy-ion reaction, namely $S+S$ at $200$ AGeV. Application of 
the relativistic (3+1)-dimensional hydrodynamic code HYLANDER \cite{udo,jan} 
yielded a successful description \cite{jan} of rapidity and transverse momentum
spectra of mesons and baryons and a quantitatively correct prediction of BEC 
of pions \cite{bernd3,alber}.

In order to get reasonable descriptions for $t(r)$, $u(r)$, $\tau(r)$ and 
$u_\perp(r)$ we have chosen the following parametrizations and parameters from 
the hydrodynamical solution mentioned above:
\ba
t(r)\:=\:\tau(r)\:=&&\left\{
\begin{array}
{r@{\quad:\quad}l}
\tau_0\:-\:\frac{1}{\alpha_0}r^2 & r \leq R_\perp\\
0 & r > R_\perp
\end{array}
\right. {\rm with}\quad \tau_0\:c/fm\:=\alpha_0\:/fm \:c\:=\:4.5\:\:,
\label{eq:parama} 
\\
u(r)\:=\:u_\perp(r)\:=&&\left\{
\begin{array}
{r@{\quad:\quad}l}
u_0 r & r \leq R_\perp\\
0 & r > R_\perp
\end{array}
\right. {\rm with}\quad u_0\:fm/c\:=\:0.13\:\:.
\label{eq:paramb} 
\ea     

The pionic chemical potential was taken to be $\mu_\pi=0.139 \:GeV$, since we 
consider the formation of a pionic Bose-condensate. The transverse radius has 
been fixed to $R_\perp=3.8\: fm$.\\

The remaining parameter is the condensate number density $n_{co}$ which 
we write in fractions of the thermal number density

\be
n_{th}\:=\:\frac{g_\pi}{(2\pi)^2} \int_0^{2\pi} k^2 dk \frac{1}
{\exp\left[\displaystyle{\frac{E\:-\:\mu_\pi}{T_f}} \right]\:-\:1}\:.
\label{eq:densth}
\ee

In Fig. 1 the results of the numerical evaluations of the single inclusive 
momentum distributions $EdN/d^3k$, the momentum-dependent chaoticities $p$ and
the Bose-Einstein correlation functions $C_2$ are shown for the spherically and
for the longitudinally expanding source. The quantity $n_{co}$ here takes four 
different values ranging from $0.0\: n_{th}$ to $1.0 \:n_{th}$.\\
It can be seen that the results for the spherical and the longitudinal 
expansion scenario look quite similar although there are some quantitative 
differences according to the momentum dependence. The Bose-condensate affects 
the plotted functions only over a limited momentum range. This is due to the 
fact that there exists a maximum velocity $u_{max}=u(R_\perp)=u_\perp(R_\perp)
=0.494 \:c$ that enters in all condensate contributions. Thus the maximum 
momentum where the Bose-condensate contributes to the single inclusive 
momentum distributions is $k_{max}=m_\pi u_{max}=68.7\: MeV/c$ and the maximum
momentum difference over which the BEC can have a reduction is twice this 
value, namely $q_{max}=137.4 \:MeV/c$.\\
Of course, theoretically the largest value $u_{max}$ could assume is the speed 
of light $c=1$. Therefore, there exists an absolute maximum value for the
momenta of particles which originate from a Bose-condensate: in case of a 
pionic Bose-condensate it would be $k_{max}=m_\pi c=139.6\: MeV/c$. 
Bose-Einstein correlation functions would be then affected over a momentum 
difference range with $q_{max}=279.2 \:MeV/c$, which would result in a 
disappearance of the sharp peak in the BEC for $S+S$, because this peak would 
be shifted to unobservably large momentum differences. It should be mentioned 
that from such a structure in the double inclusive signal one could in
principle determine the maximum flow velocity of the fluid.\\
The single inclusive momentum distributions are quite sensitive to the presence
of a Bose-conden\-sate; for the BEC this effect is even more pronounced. The 
presence of a Bose-condensate of about only $1\%$ (i.e. $\kappa \equiv 
n_{co}=0.01\: n_{th}$) results in a decrease of the intercept of the shown 
two-particle correlation functions to $1.83$ for the spherically expanding 
source and $1.86$ for the longitudinally expanding source.
In quantum optics as well as in the approach used in \cite{Fowler} the 
sensitivity of the intercept on the amount of coherence is much weaker. Thus to
get a decrease of the (``true") intercept from 2 to 1.8, a value of 
$\kappa=0.5$ is necessary. Furthermore, due to a limited value of $q_{max}$ a 
part of the tail of the two-particle correlation functions will not be 
affected by the pionic bose-condensate and a peak will therefore come into 
existence. To what extent such peaks due to an existing pionic Bose-condensate 
can be observed in experimental data depends on several factors:\\
(i) the size of the source, which determines the inverse width
of the correlation functions;\\
(ii) the maximum of accessible velocity at freeze-out of the
pionic source which determines the position of the peak;\\
(iii) deformation of the correlation function due to further 
contributions from resonance decay;\\
(iv) the averaging over acceptance regions in the experiment;\\
(v) the width of the momentum distribution in the bosonic
ground state.\\

A study of the influenece of these factors on the effects reported here is in 
preparation \cite{future}.

\section{Summary}

We have shown that the formation of a pionic Bose condensate can 
influence single inclusive spectra as well as Bose-Einstein correlation 
functions quite strongly. Among other things we find an enhancement of single 
inclusive momentum spectra at low momenta, and a reduction of the two-particle
correlation function on a limited momentum range resulting in a bumpy structure
of the BEC. The structure depends on the maximum velocity of the fluid. 
Within this treatment an explicitly momentum-dependent chaoticity
is presented for the first time due to a maximum freeze-out velocity rather 
than to geometrical effects of the source.\\
Whether these effects will survive in more realistic treatments has yet to be 
proven. However the results obtained so far confirm the interest and need to 
take into account coherence in theoretical approaches to BEC.

This work was supported by the University of
California and the Deutsche Forschungsgemeinschaft (DFG). 
B.R. Schlei acknowledges a DFG postdoctoral fellowship and R.W.
is indebted to A. Capella for the hospitality extended to him at the 
LPTHE, Univ. Paris-Sud.

\newpage
\noindent
{\Large \bf Figure Captions}\\
\begin{description}
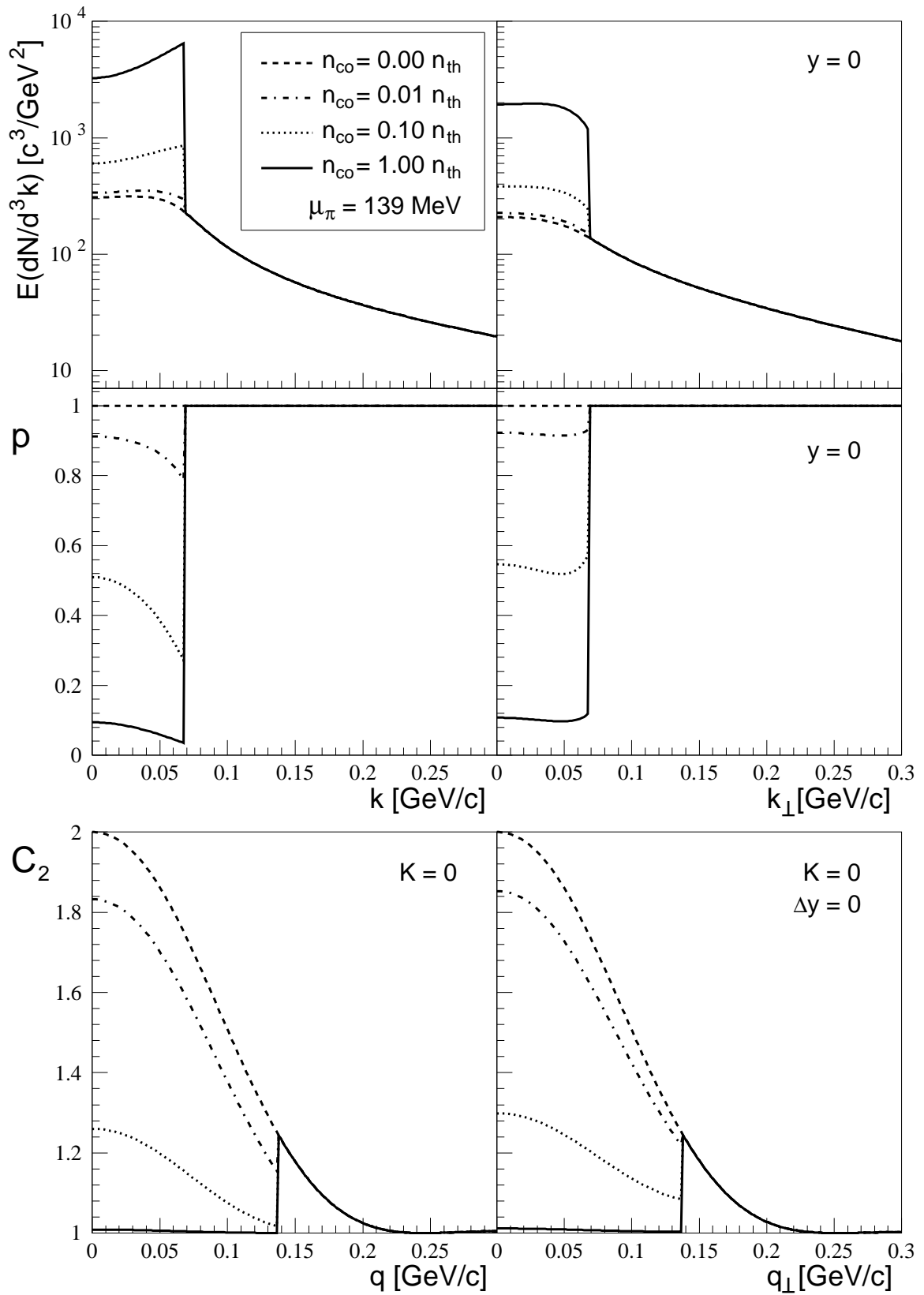
\item[Fig. 1] Single inclusive spectra, momentum-dependent
chaoticities and BE correlation functions for a spherically
and a longitudinally expanding source, respectively.
The different line styles correspond to different
condensate densities $n_{co}$ compared to the 
thermal number densities $n_{th}$. The left column of the
figure shows the results for the spherically expanding source,
whereas the right column shows the results for the longitudinally
expanding source.
\end{description}


\newpage
\begin{thebibliography}{99}
\bibitem{pbpb} S. Margetis and the NA49 Collaboration,
Nucl. Phys. \underline{A590} (1995) 355c.
\bibitem{dcc} S. Gavin, Nucl.Phys. \underline{A590} (1995) 163c. 
\bibitem{kogan} I.J. Kogan, JETP Lett. \underline{59} (1994) 307;
R.D. Amado and I.J. Kogan, Phys. Rev. \underline{D51} (1995) 190. 
\bibitem{dcc1} I.V. Andreev and R.M. Weiner, Phys.
Lett. \underline{B373} (1996) 159.
\bibitem{dan} U. Ornik, M. Pl\"umer, D. Strottman,
Phys. Lett. \underline{B314} (1993) 401.
\bibitem{Fowler} G.N. Fowler, N. Stelte, and R.M. Weiner,
Nucl. Phys. \underline{A319} (1979) 349.
\bibitem{shuryak}E.V. Shuryak, Sov.J.Nucl.Phys. \underline{18}, 667 (1974).
\bibitem{podgor} G.I. Kopylov and M.J. Podgoretsky, Sov.J.Nucl.Phys. 
\underline{18}, 336 (1974).
\bibitem{gkw} M. Gyulassy, S.K. Kauffmann, and L.W. Wilson, Phys. Rev.
\underline{C20} (1979) 2267.
\bibitem{apw}  I.V. Andreev, M. Pl\"umer, and R.M. Weiner, Phys. Rev. Lett. 
\underline{67} (1991) 3475; Int. J. Mod. Phys. \underline{A8} (1993) 4577. 
\bibitem{bernd1} B.R. Schlei, U. Ornik, M. Pl\"umer, R.M. Weiner,
Phys. Lett. \underline{B293} (1992) 275.
\bibitem{udo} U. Ornik, F. Pottag, R.M. Weiner, Phys. Rev. Lett.
\underline{63} (1989) 2641.
\bibitem{jan} J. Bolz, U. Ornik, R.M. Weiner,
Phys. Rev. \underline{C46} (1992) 2047.
\bibitem{bernd3} J. Bolz, U. Ornik, M. Pl\"umer, B.R. Schlei, R.M. Weiner,
Phys. Lett. \underline{B300} (1993) 404; Phys. Rev. \underline{D47}
(1993) 3860.
\bibitem{alber} Th. Alber et al., Phys. Rev. Lett. \underline{74} (1995)
1303; Z. Phys. \underline{C66} (1995) 77; Th. Alber for the Collaborations
NA35 and NA49, Nucl. Phys. \underline{A590} (1995) 453c.
\bibitem{future} B.R. Schlei et al., in preparation.
\end{thebibliography}
\end{document}